\documentclass{egpubl}
\usepackage{eg2026}

\ShortPresentation      % uncomment for (final) Short Conference Presentation <-- 이 부분 주석 해제 (앞의 % 제거)% \STAR                   % uncomment for STAR contribution
\usepackage[T1]{fontenc}
\usepackage{dfadobe}
\usepackage{amsfonts}
\usepackage{url}

\usepackage{egweblnk}
\usepackage{cite}  % comment out for biblatex with backend=biber
\BibtexOrBiblatex
\electronicVersion
\PrintedOrElectronic
\ifpdf \usepackage[pdftex]{graphicx} \pdfcompresslevel=9
\else \usepackage[dvips]{graphicx} \fi

\usepackage{booktabs}
\usepackage{siunitx}
\usepackage{multirow}
\usepackage{makecell}
\usepackage{graphicx}
\usepackage{subcaption}
\title{OT-UVGS: Revisiting UV Mapping for Gaussian Splatting as a Capacity Allocation Problem}

\author[Byunghyun Kim]
{
  \parbox{\textwidth}{\centering
    Byunghyun Kim\textsuperscript{1}\orcid{0009-0005-4891-3487}
  }\\
  {\parbox{\textwidth}{\centering
    \textsuperscript{1}KAIST, Republic of Korea\\
  }}
}

\begin{document}

\teaser{
    \setcounter{figure}{-1}
    \centering
    \begin{subfigure}[t]{0.16\textwidth}
        \centering
        \includegraphics[width=\linewidth]{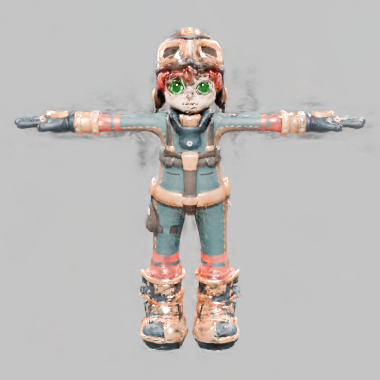}
        \caption{UVGS}
    \end{subfigure}
    \begin{subfigure}[t]{0.16\textwidth}
        \centering
        \includegraphics[width=\linewidth]{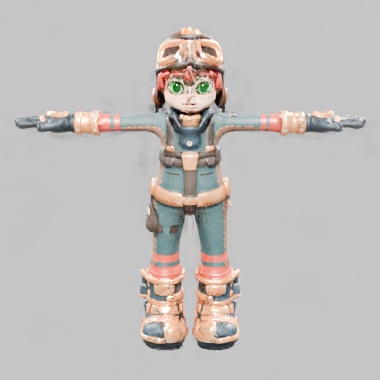}
        \caption{OT-UVGS}
    \end{subfigure}\hfill
    \begin{subfigure}[t]{0.16\textwidth}
        \centering
        \includegraphics[width=\linewidth]{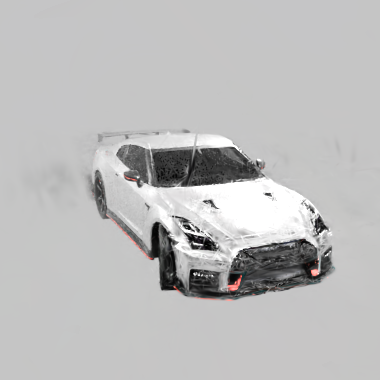}
        \caption{UVGS}
    \end{subfigure}
    \begin{subfigure}[t]{0.16\textwidth}
        \centering
        \includegraphics[width=\linewidth]{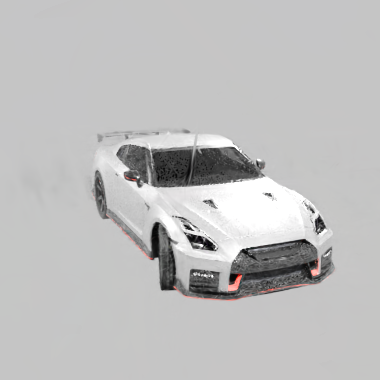}
        \caption{OT-UVGS}
    \end{subfigure}\hfill
    \begin{subfigure}[t]{0.16\textwidth}
        \centering
        \includegraphics[width=\linewidth]{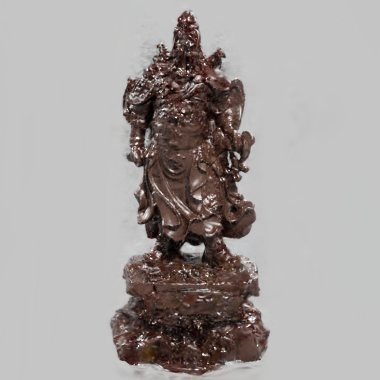}
        \caption{UVGS}
    \end{subfigure}
    \begin{subfigure}[t]{0.16\textwidth}
        \centering
        \includegraphics[width=\linewidth]{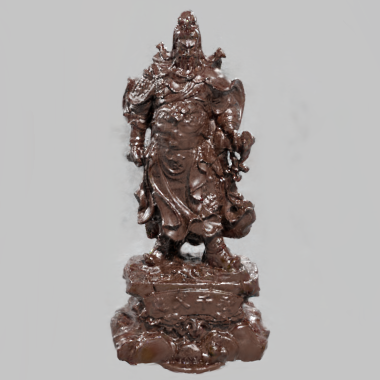}
        \caption{OT-UVGS}
    \end{subfigure}
    \captionsetup{type=figure}
    \caption{\textit{
    Qualitative comparison on three representative scenes under the same UV resolution and per-slot capacity $K{=}1$.
    Each column pair compares spherical UVGS with our OT-based mapping.
    OT-UVGS consistently reduces holes and view-dependent artifacts across scenes and views.
    }}
    \label{fig:qualitative}
    \vspace{0.5em}
}
\maketitle

\begin{abstract}
UV-parameterized Gaussian Splatting (UVGS) maps an unstructured set of 3D Gaussians to a regular UV tensor, enabling compact storage and explicit control of representation capacity.
Existing UVGS, however, uses a deterministic spherical projection to assign Gaussians to UV locations.
Because this mapping ignores the global Gaussian distribution, it often leaves many UV slots empty while causing frequent collisions in dense regions.
We reinterpret UV mapping as a capacity-allocation problem under a fixed UV budget and propose \emph{OT-UVGS}, a lightweight, separable one-dimensional optimal-transport-inspired mapping that globally couples assignments while preserving the original UVGS representation.
The method is implemented with rank-based sorting, has $O(N\log N)$ complexity for $N$ Gaussians, and can be used as a drop-in replacement for spherical UVGS.
Across 184 object-centric scenes and the Mip-NeRF dataset, OT-UVGS consistently improves peak signal-to-noise ratio (PSNR), structural similarity (SSIM), and Learned Perceptual Image Patch Similarity (LPIPS) under the same UV resolution and per-slot capacity ($K{=}1$).
These gains are accompanied by substantially better UV utilization, including higher non-empty slot ratios, fewer collisions, and higher Gaussian retention.
Our results show that revisiting the mapping alone can unlock a significant fraction of the latent capacity of UVGS.



\end{abstract}

%-------------------------------------------------------------------------
\section{Introduction}
\label{sec:intro}

3D Gaussian Splatting (3DGS)~\cite{KerblKKLD23} has emerged as an effective representation for novel view synthesis, combining high visual fidelity with efficient optimization and real-time rendering.
However, unstructured Gaussian sets still pose challenges for memory locality, compression, and interoperability with standard 2D and 3D learning pipelines.
This has motivated recent work on \emph{structuring} Gaussian parameters into regular representations with explicit capacity control~\cite{XuHLSZ24_TextureGS,JiangLZLZ24_UVGaussians,LuYXXWLBD23_ScaffoldGS}.

UV-parameterized Gaussian Splatting (UVGS)~\cite{RaiWJ*25} addresses this goal by mapping an unstructured Gaussian set $\mathcal{G}$ to a regular UV tensor
$\Phi(\mathcal{G}) \in \mathbb{R}^{H \times W \times K \times C}$,
where $H$ and $W$ denote the UV height and width, $K$ denotes the maximum number of Gaussians stored per UV slot, and $C$ denotes the number of channels used to encode Gaussian attributes.
This formulation provides a compact 2D layout with a fixed representation budget and enables direct reuse of mature 2D and 3D modeling techniques.
As a result, UVGS offers a promising abstraction for structured Gaussian representations.

Despite this promise, we identify a key bottleneck in existing UVGS~\cite{RaiWJ*25}: the UV mapping itself.
Current approaches assign Gaussians to UV coordinates via a deterministic spherical projection.
Because this pointwise mapping ignores the global distribution of Gaussians, it often under-utilizes the available UV budget: large portions of the UV grid remain empty, whereas dense regions suffer from repeated collisions.
Consequently, increasing the UV resolution does not reliably translate into higher effective capacity, and performance often depends on using a large $K$ to compensate.

We reinterpret UV mapping as a capacity-allocation problem under a fixed UV budget.
From this viewpoint, the goal is to distribute Gaussians so that UV slots are used as uniformly as possible, which naturally suggests matching the empirical Gaussian distribution to a uniform target measure over the UV domain.

Based on this insight, we propose OT-UVGS, a lightweight separable 1D OT-inspired mapping implemented via rank-based sorting with $O(N\log N)$ complexity.
OT-UVGS preserves the original UVGS representation and modifies only the mapping strategy, making it a drop-in replacement for spherical UVGS.

Experiments show that changing only the mapping yields consistent gains under identical UV resolution and $K$.
OT-UVGS improves PSNR and SSIM, reduces LPIPS, and substantially weakens the dependence on large per-slot capacity.
These gains hold across 184 object-centric scenes and full-scene evaluations on the Mip-NeRF dataset~\cite{BarronMMTPS21_MipNeRF}, demonstrating that the proposed mapping generalizes beyond isolated-object settings.

\noindent\textbf{Contributions.}
(1) We identify UV mapping as a primary bottleneck in UVGS and reinterpret it as a capacity-allocation problem.
(2) We propose a rank-based separable 1D OT-inspired mapping that improves UV slot utilization with $O(N\log N)$ complexity.
(3) We demonstrate consistent improvements in rendering quality and reduced dependence on large per-slot capacity under identical $(H,W,K)$.

%-------------------------------------------------------------------------
\section{Background}
\label{sec:background}

\noindent\textbf{UVGS recap.}
Let $\mathcal{G} = \{ g_i \}_{i=1}^{N}$ denote a set of $N$ 3D Gaussians, where each Gaussian
$g_i = (\mathbf{x}_i, \mathbf{s}_i, \mathbf{r}_i, \alpha_i, \mathbf{c}_i)$
stores its position $\mathbf{x}_i \in \mathbb{R}^{3}$, scale $\mathbf{s}_i$, rotation $\mathbf{r}_i$, opacity $\alpha_i$, and appearance coefficients $\mathbf{c}_i$.
UVGS~\cite{RaiWJ*25} maps this unstructured set to a structured UV tensor
\begin{equation}
\Phi(\mathcal{G}) \in \mathbb{R}^{H \times W \times K \times C},
\end{equation}
where $H$ and $W$ denote the UV grid height and width, $K$ is the maximum number of Gaussians stored in each UV slot, and $C$ is the number of channels used to encode the retained Gaussian attributes.
This parameterization enables compact storage, regular memory access, and explicit control of representation capacity.

\noindent\textbf{Limitation of spherical mapping.}
Existing UVGS assigns Gaussians to UV coordinates using a deterministic spherical projection~\cite{RaiWJ*25,XuHLSZ24_TextureGS} computed from angular coordinates.
Because each Gaussian is mapped independently, the method ignores the global distribution of the Gaussian set.
The result is poor capacity allocation: many UV slots remain empty, while dense angular regions suffer from repeated collisions.
Consequently, increasing the UV resolution does not necessarily increase the effective capacity, and performance often depends on using a larger $K$ to absorb these collisions.

\noindent\textbf{OT perspective on UV mapping.}
We instead view UV mapping as a capacity-allocation problem under a fixed budget $HWK$.
The goal is to distribute Gaussians so that the available UV slots are used as uniformly as possible.
Optimal transport (OT) provides a natural framework for matching an empirical distribution to a target distribution while introducing global coupling between assignments~\cite{PeyreCuturi20_COT,GenestBNC25_BSPOT}.
Our method adopts this perspective but uses a lightweight separable construction that requires only sorting rather than a general discrete OT solver.

\noindent\textbf{Angular histogram equalization (HE).}
As a baseline, we consider angular histogram equalization (HE), which independently equalizes the marginal distributions of $\theta$ and $\phi$ without global coupling.

\begin{table}[t]
\centering
\caption{\textit{Overall object-centric comparison on 184 scenes. We report mean $\pm$ standard deviation over scenes.}}
\label{tab:main}
\setlength{\tabcolsep}{6pt}
\begin{tabular}{lccc}
\toprule
Method & PSNR$\uparrow$ & SSIM$\uparrow$ & LPIPS$\downarrow$ \\
\midrule
UVGS
& 35.53 $\pm$ 7.36
& 0.9446 $\pm$ 0.0665
& 0.0311 $\pm$ 0.0248 \\
HE
& 38.03 $\pm$ 6.84
& 0.9786 $\pm$ 0.0256
& 0.0178 $\pm$ 0.0181 \\
Ours
& \textbf{41.20} $\pm$ 6.35
& \textbf{0.9879} $\pm$ 0.0101
& \textbf{0.0112} $\pm$ 0.0113 \\
\bottomrule
\end{tabular}
\end{table}

\begin{table}[t]
\centering
\caption{\textit{Capacity-utilization analysis under the same UV resolution and per-slot capacity $K{=}1$. Gaussian retention denotes the fraction of input Gaussians kept after per-slot top-$K$ truncation. Values are mean $\pm$ standard deviation over scenes.}}
\label{tab:utilization}
\small
\setlength{\tabcolsep}{3pt}
\begin{tabular}{lccc}
\toprule
Method & Non-empty UV (\%)$\uparrow$ & Collision rate$\downarrow$ & Gaussian retention (\%)$\uparrow$ \\
\midrule
UVGS & 22.83 $\pm$ 6.49 & 0.355 $\pm$ 0.090 & 64.89 $\pm$ 17.43 \\
HE & 28.20 $\pm$ 5.36 & 0.240 $\pm$ 0.043 & 77.34 $\pm$ 5.31 \\
Ours & \textbf{30.30} $\pm$ 4.29 & \textbf{0.226} $\pm$ 0.041 & \textbf{83.29} $\pm$ 9.46 \\
\bottomrule
\end{tabular}
\end{table}

\begin{figure}[t]
    \centering

    \begin{subfigure}[t]{0.48\columnwidth}
        \centering
        \includegraphics[width=\linewidth]{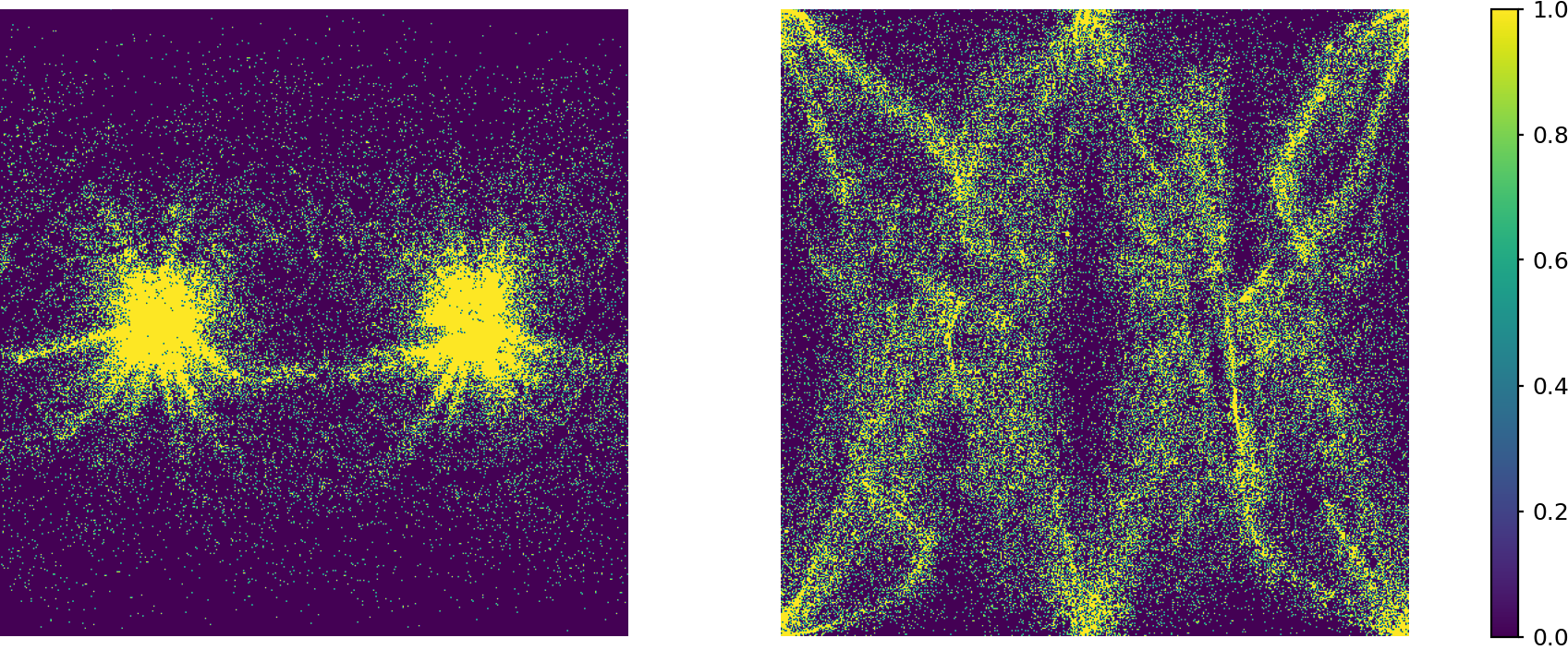}
        \caption{Scene A}
    \end{subfigure}\hfill
    \begin{subfigure}[t]{0.48\columnwidth}
        \centering
        \includegraphics[width=\linewidth]{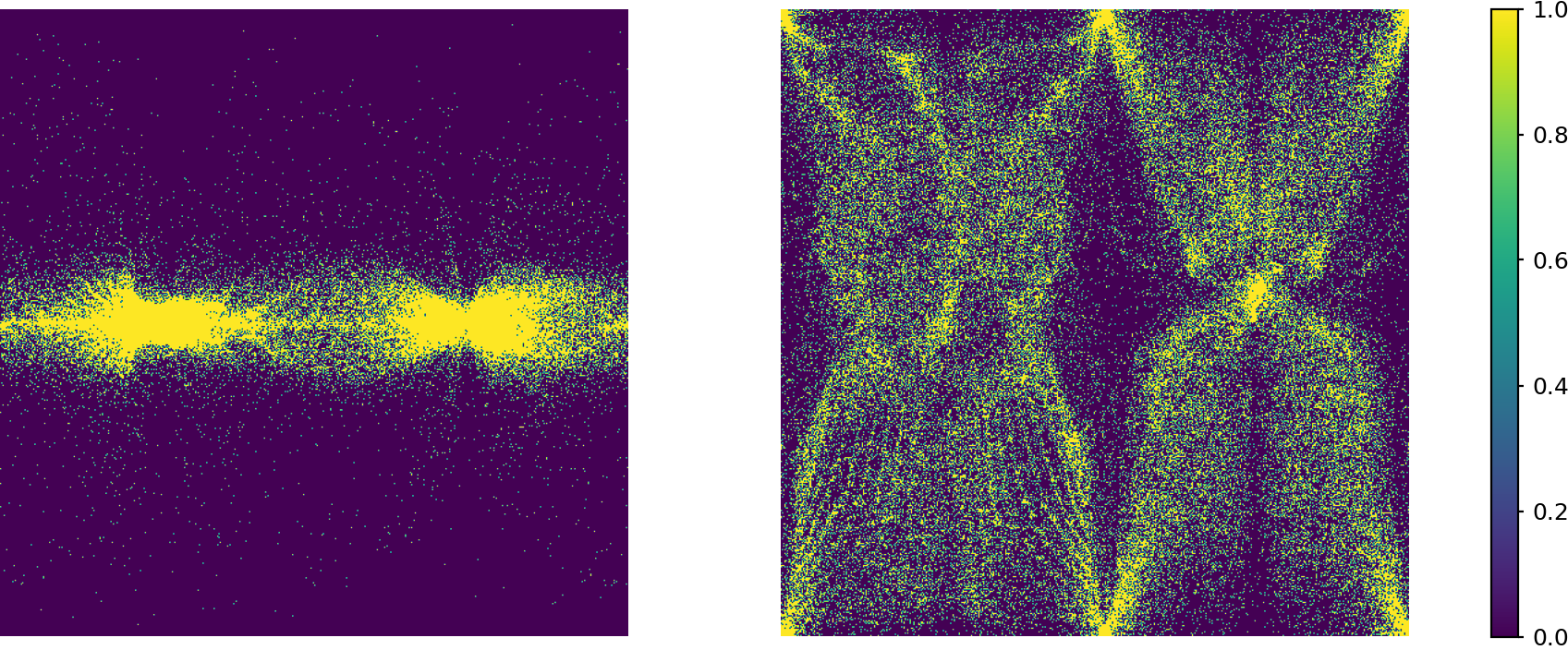}
        \caption{Scene B}
    \end{subfigure}

    \vspace{0.4em}

    \begin{subfigure}[t]{0.48\columnwidth}
        \centering
        \includegraphics[width=\linewidth]{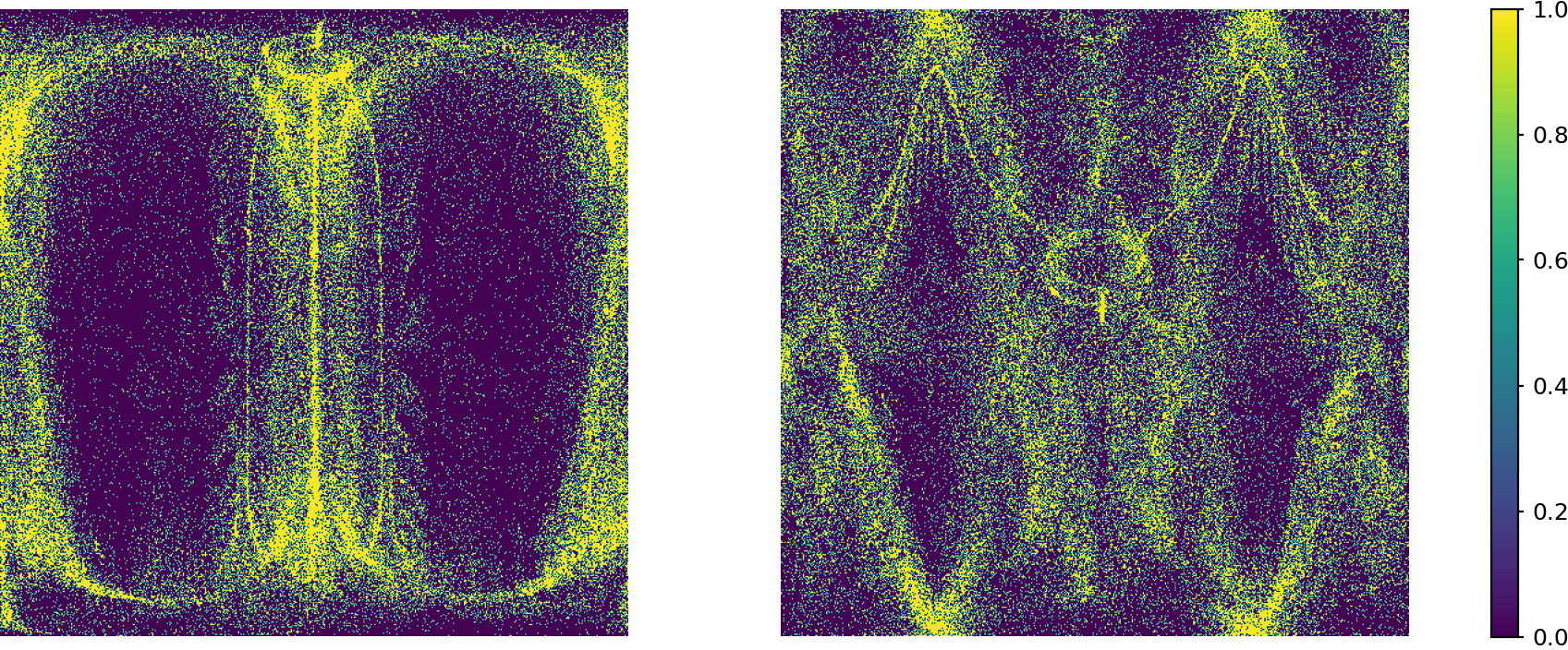}
        \caption{Scene C}
    \end{subfigure}\hfill
    \begin{subfigure}[t]{0.48\columnwidth}
        \centering
        \includegraphics[width=\linewidth]{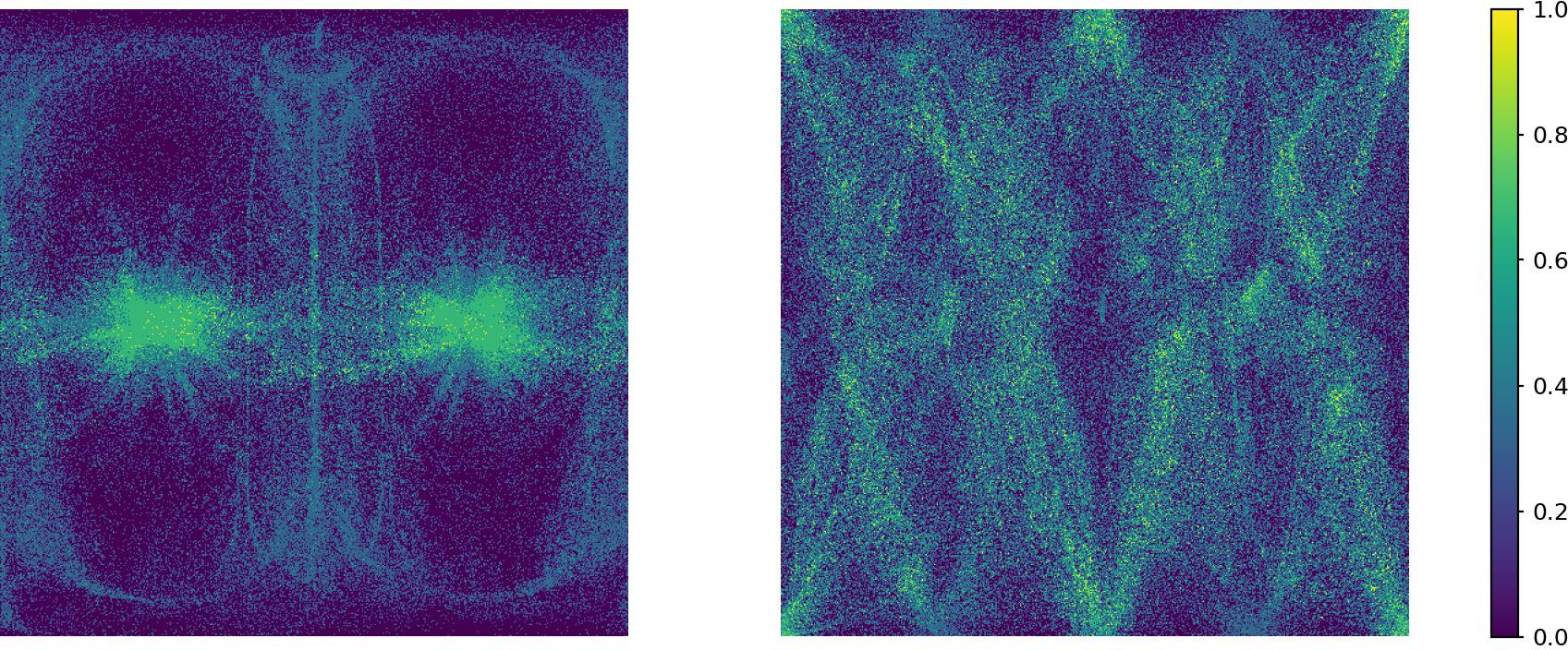}
        \caption{All scenes}
    \end{subfigure}

    \caption{\textit{
    UV occupancy heatmaps comparing spherical UVGS (left in each pair) and OT-UVGS (right in each pair) under the same UV resolution and $K{=}1$.
    Spherical mapping exhibits strong angular concentration and large empty regions, whereas OT-UVGS distributes Gaussians more uniformly across the UV domain.
    }}
    \label{fig:uv_occupancy}
\end{figure}

\section{Method}
\label{sec:method}

\noindent\textbf{Problem setting.}
Given the Gaussian set $\mathcal{G}$, the UV grid size $(H, W)$, and the per-slot capacity $K$, our goal is to improve the assignment stage while keeping the UVGS tensor layout, stored attributes, renderer, and training protocol unchanged.
OT-UVGS therefore replaces only the mapping from Gaussians to UV coordinates.

\noindent\textbf{Directional coordinates.}
For each Gaussian position $\mathbf{x}_i$, we compute the normalized direction
\begin{equation}
\hat{\mathbf{x}}_i = (\hat{x}_i, \hat{y}_i, \hat{z}_i) = \frac{\mathbf{x}_i}{\|\mathbf{x}_i\|_2}.
\end{equation}
We then convert this unit vector to spherical angles
\begin{equation}
\theta_i = \mathrm{atan2}(\hat{y}_i, \hat{x}_i), \qquad
\phi_i = \arccos(\hat{z}_i),
\end{equation}
where $\theta_i \in [-\pi,\pi)$ is the azimuth and $\phi_i \in [0,\pi]$ is the polar angle.

\noindent\textbf{Uniform capacity measure.}
We treat each Gaussian as one unit of representation capacity and define the empirical discrete measure
\begin{equation}
\mu = \frac{1}{N} \sum_{i=1}^{N} \delta_{g_i},
\end{equation}
where $\delta_{g_i}$ denotes a Dirac mass at $g_i$.
The target is a uniform allocation over the UV domain, because our objective is not to preserve physical mass but to use the fixed representation budget as evenly as possible.
For this reason, we do not weight Gaussians by opacity or scale.

\noindent\textbf{Rank-based separable OT mapping.}
We compute the empirical cumulative distribution functions of the two angular coordinates by sorting:
\begin{equation}
u_i = \frac{\mathrm{rank}(\theta_i)}{N}, \qquad
v_i = \frac{\mathrm{rank}(\phi_i)}{N},
\end{equation}
where $\mathrm{rank}(\cdot) \in \{1,\ldots,N\}$ is the ascending rank among all Gaussians.
This procedure maps the empirical marginals of $\theta$ and $\phi$ to uniform marginals on $[0,1]$.
Although the construction is separable rather than a full 2D OT solve, it is globally coupled in the sense that every assignment depends on the ordering of the entire Gaussian set.
In contrast to general discrete OT solvers~\cite{PeyreCuturi20_COT,GenestBNC25_BSPOT}, the resulting map is obtained with simple sorting and therefore retains $O(N \log N)$ complexity.

\noindent\textbf{UV discretization.}
The continuous coordinates $(u_i, v_i)$ are converted to integer UV indices $(\tilde{u}_i, \tilde{v}_i)$ via
\begin{equation}
\tilde{u}_i =
\min\!\left(W-1,\;\left\lfloor u_i W \right\rfloor\right), \qquad
\tilde{v}_i =
\min\!\left(H-1,\;\left\lfloor v_i H \right\rfloor\right).
\end{equation}
Each Gaussian is then assigned to the UV slot $(\tilde{v}_i, \tilde{u}_i)$.

\noindent\textbf{Top-$K$ retention.}
If more than $K$ Gaussians are assigned to the same slot, we retain the $K$ Gaussians with the largest opacity $\alpha_i$.
Because OT-UVGS distributes Gaussians more evenly across the UV plane, strong performance is often obtained already at $K{=}1$.

\noindent\textbf{Properties.}
OT-UVGS introduces global coupling through rank statistics, improves UV slot utilization, and can be inserted into UVGS as a drop-in replacement.
Since only the mapping changes, any quality difference directly reflects the effect of capacity allocation under the same nominal budget.

\begin{figure}[t]
    \centering
    \begin{subfigure}[t]{0.48\columnwidth}
        \centering
        \hspace*{-0.05\linewidth}
        \includegraphics[width=1.10\linewidth]{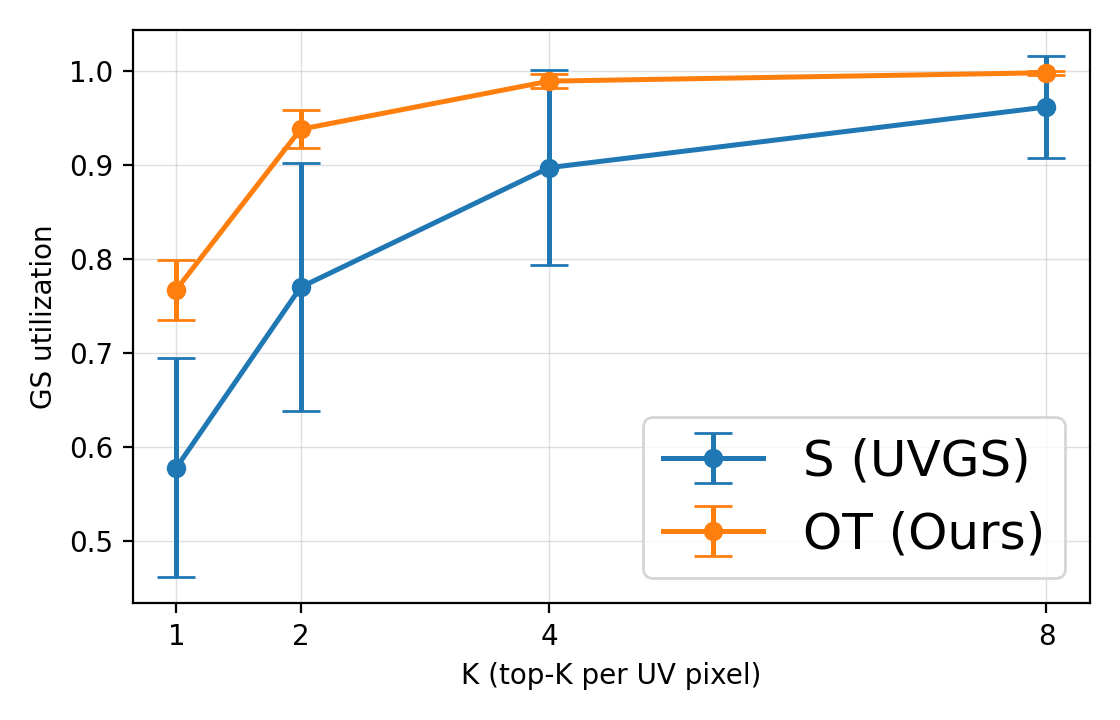}
        \caption{Gaussian retention vs.\ $K$}
        \label{fig:ksweep_util}
    \end{subfigure}
    \hspace{0.02\columnwidth}
    \begin{subfigure}[t]{0.48\columnwidth}
        \centering
        \hspace*{-0.05\linewidth}
        \includegraphics[width=1.10\linewidth]{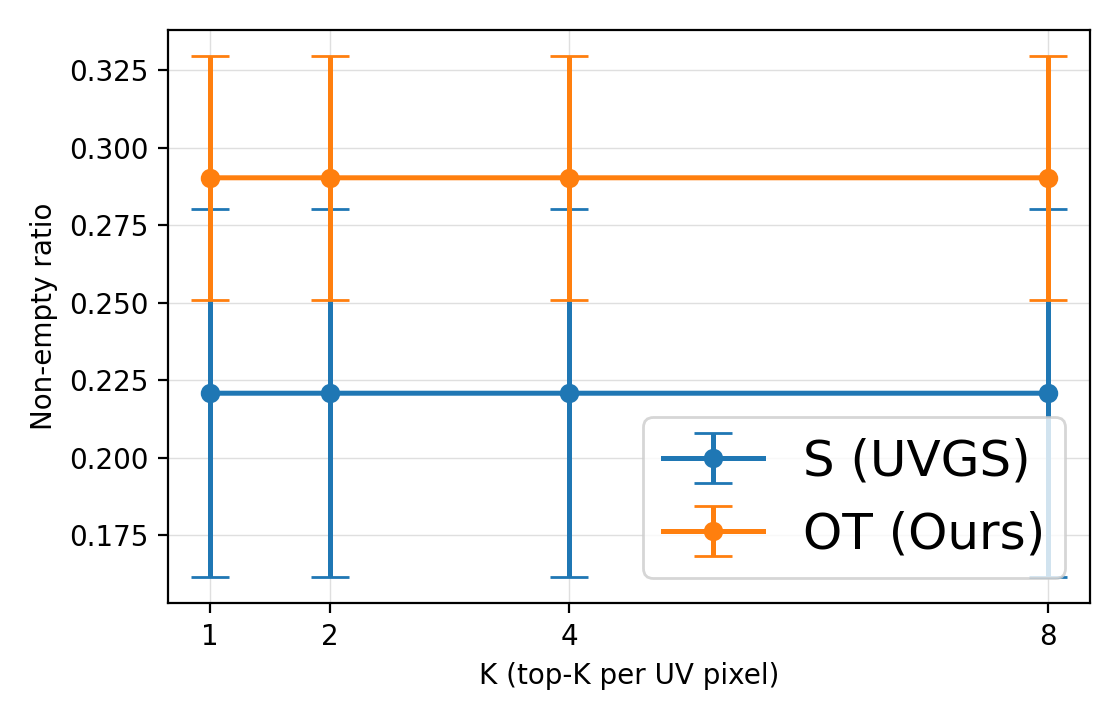}
        \caption{Non-empty UV ratio vs.\ $K$}
        \label{fig:ksweep_nonempty}
    \end{subfigure}

    \caption{\textit{Sensitivity to the per-slot capacity $K$. OT-UVGS maintains higher Gaussian retention and higher non-empty UV ratios, especially at small $K$.}}
    \label{fig:k_sweep}
\end{figure}

\section{Results}
\label{sec:results}

\noindent\textbf{Setup.}
We conduct object-centric experiments on 184 scenes sampled from Objaverse~\cite{DeitkeEtAl23_Objaverse}, following the UVGS evaluation protocol~\cite{RaiWJ*25}.
Each scene is evaluated on 48 held-out views.
We compare three mappings under the same Gaussian set, UV resolution, renderer, and training protocol: the original spherical UVGS mapping (UVGS), angular histogram equalization (HE), and OT-UVGS (OT).

\noindent\textbf{UV configuration.}
Unless otherwise stated, we use a UV resolution of $(H,W)=(512,512)$ and per-slot capacity $K{=}1$ to expose capacity under-utilization directly.
No scene-specific tuning is performed.

\noindent\textbf{Metrics.}
We report peak signal-to-noise ratio (PSNR), structural similarity (SSIM), and Learned Perceptual Image Patch Similarity (LPIPS) averaged over views.
To quantify effective capacity utilization, we additionally measure (i) the non-empty UV ratio, the fraction of UV slots that receive at least one Gaussian; (ii) the collision rate, the fraction of non-empty UV slots that receive more than one Gaussian before top-$K$ truncation; and (iii) Gaussian retention, the fraction of Gaussians kept after top-$K$ truncation.

\subsection{Object-Centric Results}
\label{sec:object_results}

Table~\ref{tab:main} reports the overall object-centric results.
Under the same UV resolution and $K{=}1$, OT-UVGS outperforms both spherical UVGS and HE on all three rendering metrics.
The gains are substantial for both fidelity-oriented metrics (PSNR and SSIM) and perceptual quality (LPIPS), and the lower standard deviations indicate that the improvement is consistent across scenes.

Table~\ref{tab:utilization} explains these gains.
OT-UVGS yields a higher non-empty UV ratio, a lower collision rate, and higher Gaussian retention under the same budget.
This shows that better capacity allocation is the main driver of the quality improvement.

\subsection{UV Slot Utilization Analysis}
\label{sec:occupancy}

To visualize the effect of the mapping, Figure~\ref{fig:uv_occupancy} shows UV occupancy heatmaps.
Spherical mapping concentrates many Gaussians in limited angular regions and leaves large empty areas elsewhere.
OT-UVGS distributes assignments more evenly across the UV plane, which directly increases the usable fraction of the fixed UV budget.

\subsection{Sensitivity to Per-Slot Capacity $K$}
\label{sec:k_sensitivity}

Figure~\ref{fig:k_sweep} analyzes sensitivity to the per-slot capacity $K$.
Spherical UVGS relies heavily on larger $K$ values to recover from collision-heavy assignments.
OT-UVGS maintains higher Gaussian retention and non-empty UV ratios even when $K$ is small, showing that improved mapping reduces the need for per-slot over-parameterization.
OT-UVGS also exhibits more stable behavior across scenes, which is desirable for memory-efficient structured representations.

\subsection{Full-Scene Generalization}
\label{sec:fullscene}

\begin{figure}[t]
    \centering
    \includegraphics[width=0.49\columnwidth]{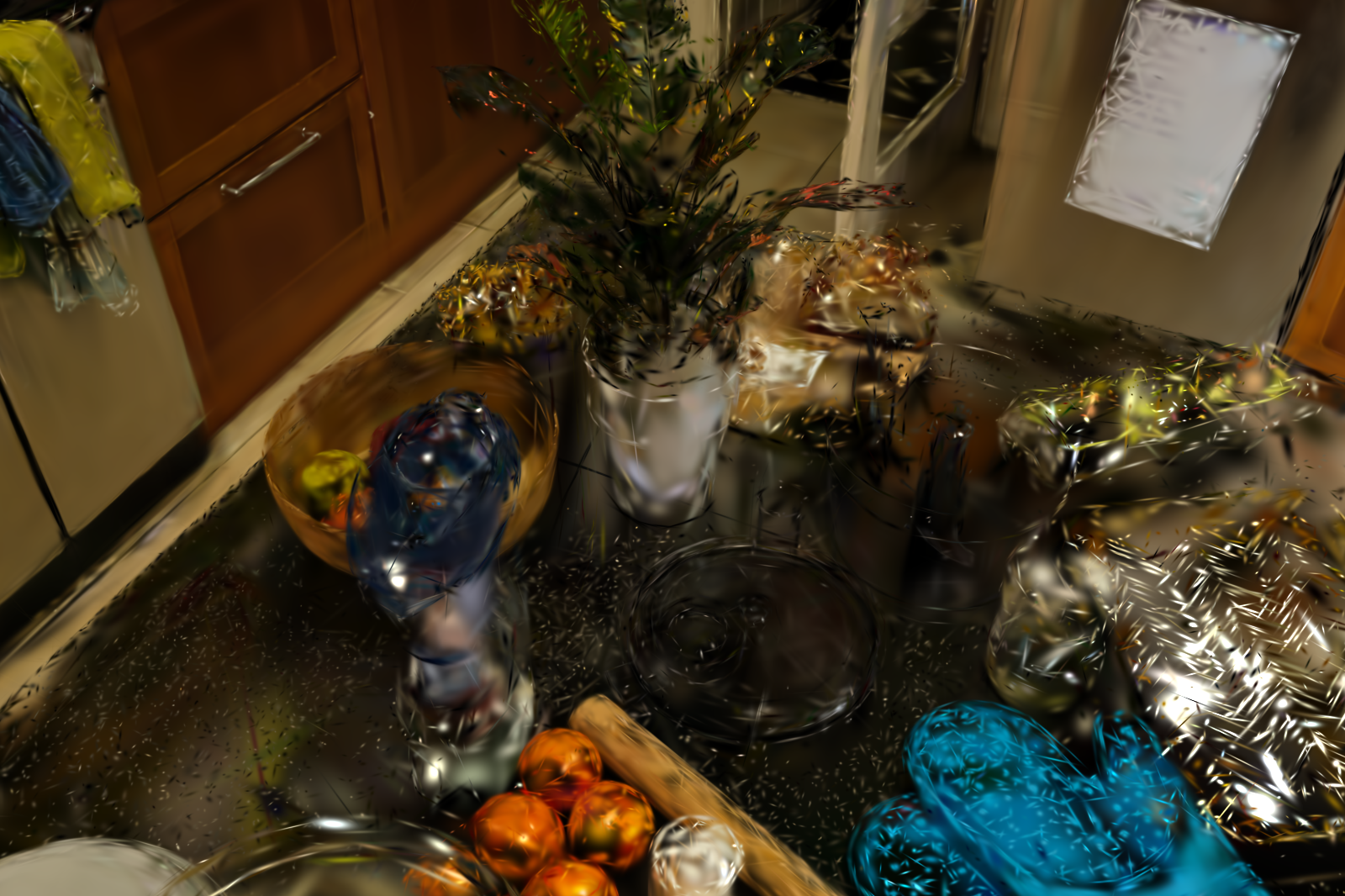}\hfill
    \includegraphics[width=0.49\columnwidth]{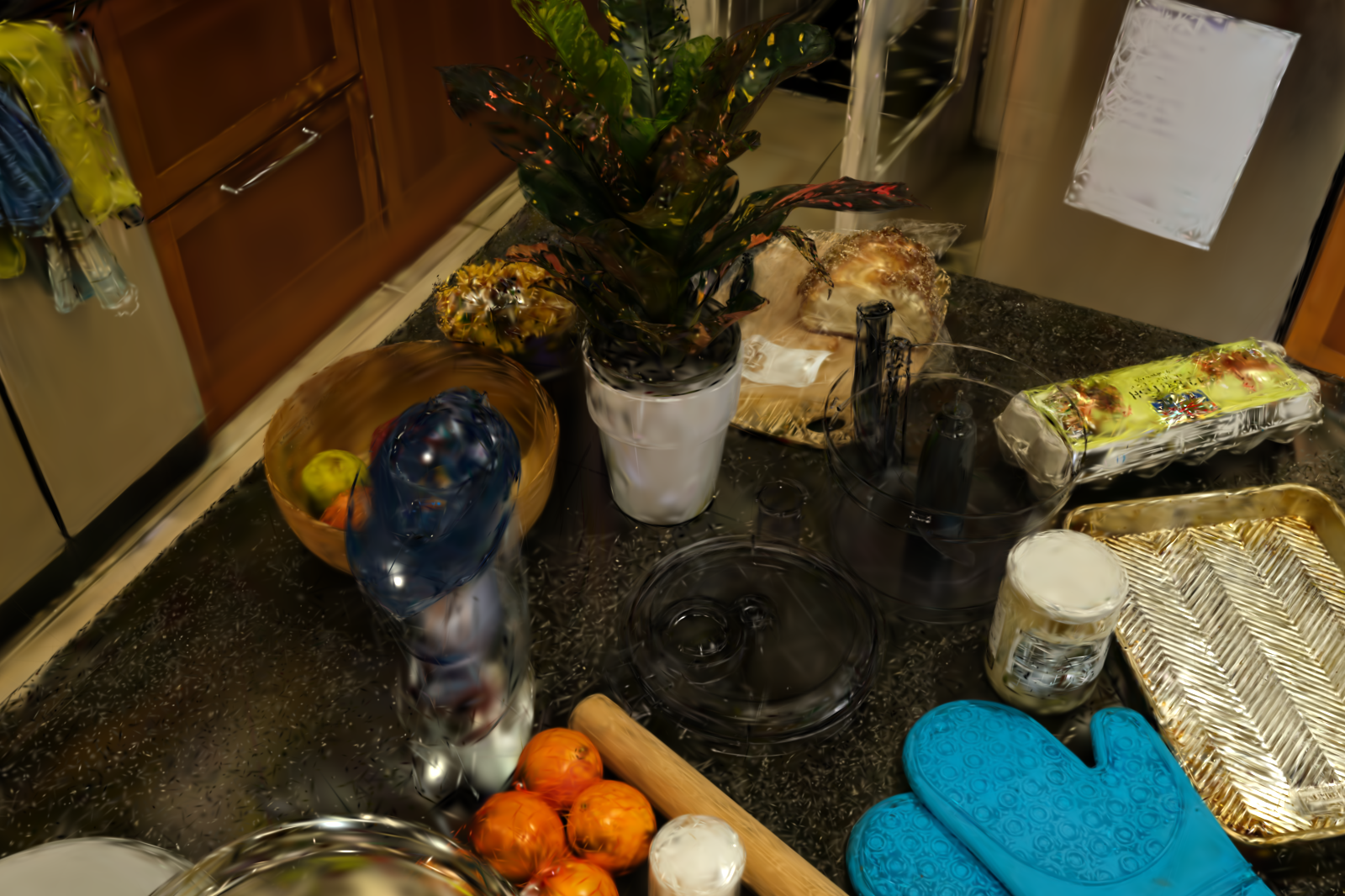}\\[3pt]
    \includegraphics[width=0.49\columnwidth]{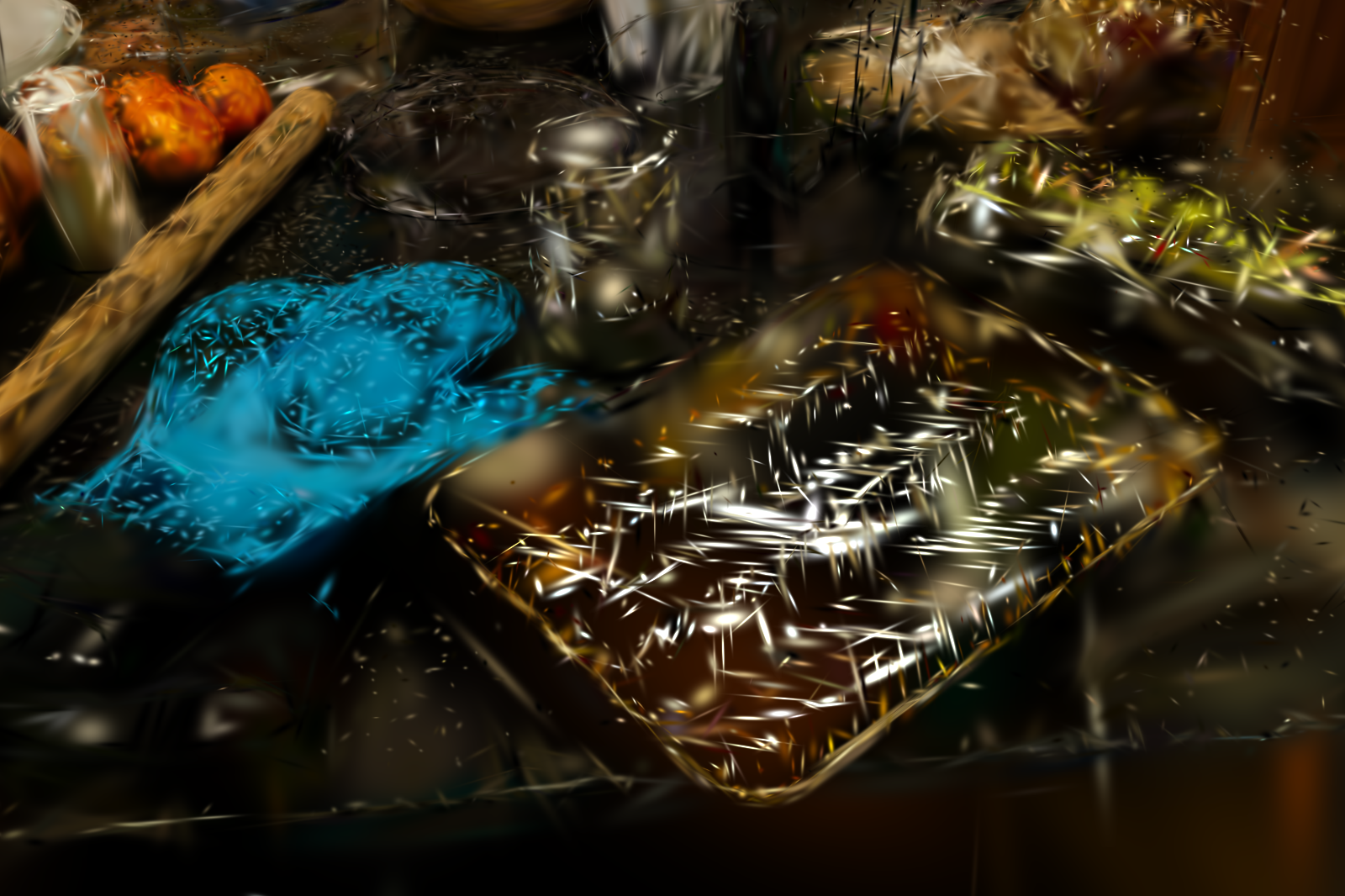}\hfill
    \includegraphics[width=0.49\columnwidth]{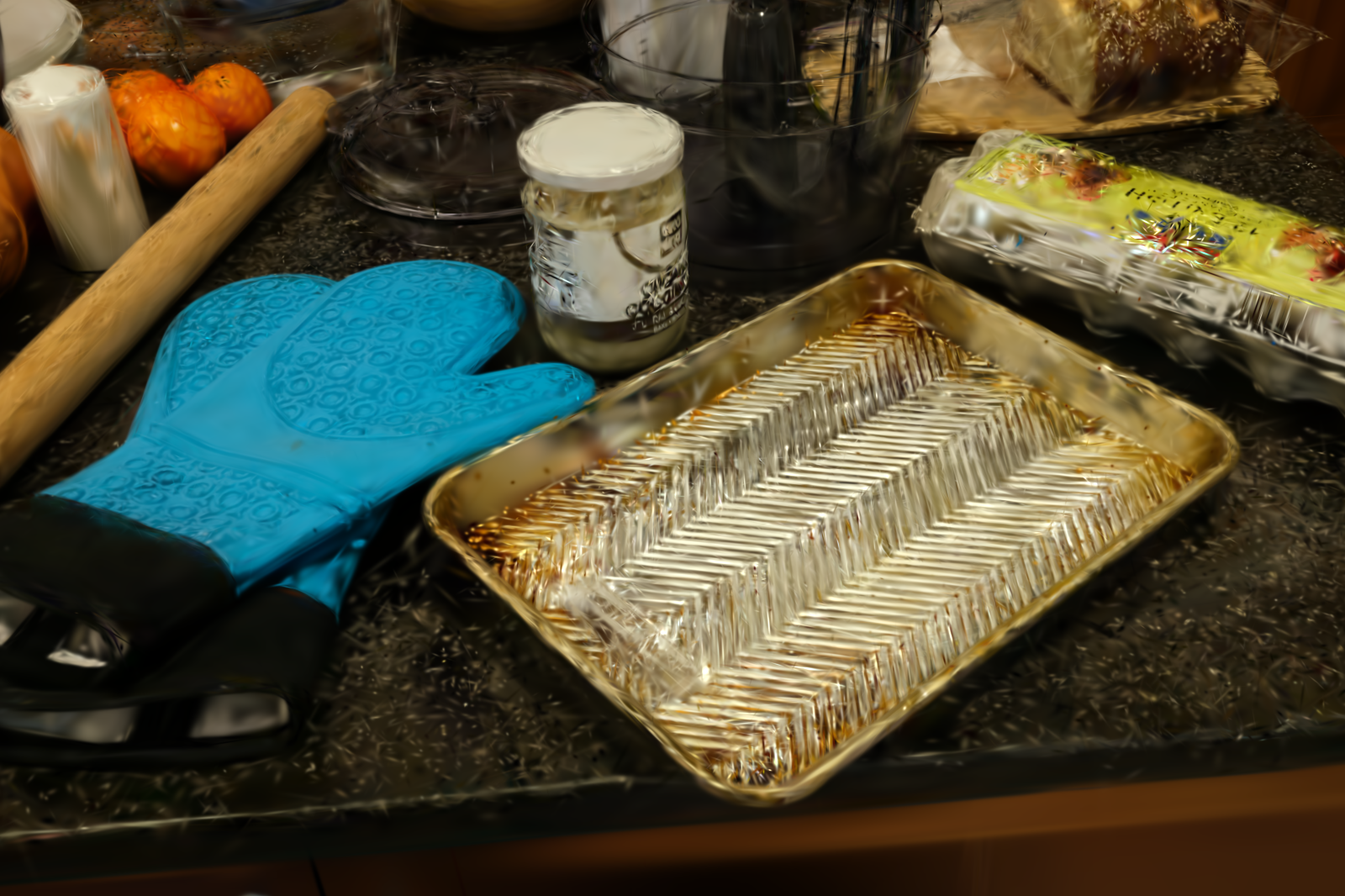}

    \caption{
    Full-scene qualitative comparison on a Mip-NeRF scene.
    Left column: spherical UVGS.
    Right column: OT-UVGS.
    OT-UVGS reduces missing regions and improves structural consistency across views under the same UV budget.
    }
    \label{fig:fullscene_qual}
\end{figure}

To evaluate generalization beyond object-centric scenes, we additionally test on the Mip-NeRF dataset~\cite{BarronMMTPS21_MipNeRF}.
Table~\ref{tab:fullscene} and Figure~\ref{fig:fullscene_qual} show that OT-UVGS also improves rendering quality in full-scene settings under the same UV budget and $K{=}1$.

Although this evaluation is preliminary, it suggests that capacity-aware mapping is not limited to isolated objects.

\begin{table}[t]
\centering
\caption{\textit{Full-scene evaluation on the Mip-NeRF dataset under the same UV budget and $K{=}1$. OT-UVGS consistently outperforms both spherical mapping and HE, showing that better capacity allocation leads to better rendering quality.}}
\label{tab:fullscene}
\small
\setlength{\tabcolsep}{6pt}
\begin{tabular}{lccc}
\toprule
Method & PSNR$\uparrow$ & SSIM$\uparrow$ & LPIPS$\downarrow$ \\
\midrule
UVGS & 14.36 $\pm$ 3.32 & 0.5612 $\pm$ 0.2124 & 0.4271 $\pm$ 0.0779 \\
HE & 15.39 $\pm$ 4.06 & 0.6818 $\pm$ 0.2373 & 0.3660 $\pm$ 0.1161 \\
Ours & \textbf{16.78} $\pm$ 3.96 & \textbf{0.7019} $\pm$ 0.2262 & \textbf{0.3450} $\pm$ 0.1150 \\
\bottomrule
\end{tabular}
\end{table}

%-------------------------------------------------------------------------
\section{Conclusion}
\label{sec:conclusion}
We identified UV mapping as a key bottleneck in UV-parameterized Gaussian Splatting and reframed it as a capacity-allocation problem under a fixed UV budget.
By introducing a lightweight rank-based separable OT mapping, we improve UV slot utilization and rendering quality without changing the underlying UVGS representation or increasing the nominal model capacity.

More broadly, our results suggest that UV mapping should be treated not merely as geometric preprocessing, but as a representation-design choice that determines how finite capacity is distributed.
Because OT-UVGS preserves the structured UV layout of UVGS, it remains compatible with existing 2D and 3D models and does not require changes to downstream rendering or learning pipelines.

Future work includes task-adaptive or learned transport objectives that tie capacity allocation to downstream supervision, as well as extensions of capacity-aware mapping to other structured splatting settings such as 2D Gaussian representations, hybrid mesh--Gaussian models, and multi-resolution layouts.
\bibliographystyle{eg-alpha-doi}
\bibliography{refs}

\end{document}